\documentclass[preprints,accept,pdftex,moreauthors]{mdpi} 

\firstpage{1} 
\pubvolume{1}
\issuenum{1}
\articlenumber{0}
\pubyear{2023}
\copyrightyear{2023}
\hreflink{https://doi.org/}

\Title{Low Noise Opto-Electro-Mechanical Modulator for RF-to-Optical Transduction in Quantum Communications}

\TitleCitation{Low Noise Opto-Electro-Mechanical Modulator for RF-to-Optical Transduction in Quantum Communications}

\Author{Michele Bonaldi $^{1,2}$, Antonio Borrielli $^{1,2}$,  Giovanni Di Giuseppe $^{3,4}$, Nicola Malossi $^{3,4}$, Bruno Morana $^{5}$, Riccardo Natali $^{3,4}$, Paolo Piergentili $^{3,4}$, Pasqualina Maria Sarro $^{5}$, Enrico Serra $^{2,5}$ and David Vitali $^{3,4,6,}$*}

\AuthorNames{Michele Bonaldi, Antonio Borrielli,  Giovanni Di Giuseppe, Nicola Malossi, Bruno Morana, Riccardo Natali, Paolo Piergentili, Pasqualina Maria Sarro, Enrico Serra and David Vitali}

\AuthorCitation{Bonaldi, M.; Borrielli, A.; Di Giuseppe, G.; Malossi, N.; Morana, B.; Natali, R.; Piergentili, P.; Sarro, P.M.; Serra, E.; Vitali, D.}

\address{$^{1}$ \quad Institute of Materials for Electronics and Magnetism, Nanoscience-Trento-FBK  Division, 38123 Povo, TN, Italy; mbonaldi@fbk.eu (M.B.); borrielli@fbk.eu (A.B.)
\\
$^{2}$ \quad Istituto Nazionale di Fisica Nucleare, TIFPA, 38123 Povo, TN, Italy; enrico.serra@tifpa.infn.it (E.S.)\\
$^{3}$ \quad Physics Division, School of Science and Technology, University of Camerino, 62032 Camerino, MC, Italy; gianni.digiuseppe@unicam.it (G.D.G.); nicola.malossi@unicam.it (N.M.); riccardo.natali@unicam.it (R.N.); paolo.piergentili@unicam.it (P.P.)\\
$^{4}$ \quad INFN, Sezione di Perugia, 06123 Perugia, PG, Italy\\
$^{5}$ \quad Department of Microelectronics and Computer Engineering, ECTM, Delft University of Technology, Feldmanweg 17, 2628 CT Delft, The Netherlands; b.morana@tudelft.nl (B.M.); p.m.sarro@tudelft.nl (P.M.S.)\\
$^{6}$ \quad CNR-INO, L.go Enrico Fermi 6, 50125 Firenze, FI, Italy \\
}

\corres{Correspondence: david.vitali@unicam.it; Tel.:+39-0737-402540}

\abstract{In this work, we present an Opto-Electro-Mechanical Modulator (OEMM) for RF-to-optical transduction realized via an ultra-coherent nanomembrane resonator capacitively coupled to an rf injection circuit
made of a microfabricated read-out able to improve the electro-optomechanical interaction. This device configuration can be embedded in a Fabry--Perot cavity for electromagnetic cooling of the LC circuit in a dilution refrigerator exploiting the opto-electro-mechanical interaction. 
To this aim, an optically measured steady-state frequency shift of 380 Hz was seen with a polarization voltage of 30 V and a $Q$-factor of the assembled device above $10^6$ at room temperature. The rf-sputtered titanium nitride layer can be made superconductive to develop efficient quantum~transducers.
}


\begin{document}
\section{Introduction}
Quantum transduction refers to the process of converting one form of energy to another at the single excitation level,  and~it represents a key ingredient in quantum technologies. Major interest is currently placed on the coherent conversion between optical and microwave/radiofrequency (mw/rf) photons~\cite{Chu2020,Lauk2020}, the~optical domain being ideal for reliable long-range communications through optical fibers or in free space, while the lower frequency band is particularly suitable for high-fidelity local quantum operations using superconducting and other solid state processors. This will allow a global quantum Internet or distributed quantum tasks including computing or sensing~\cite{Kimble2008,Pirandola2018}. Moreover, quantum transduction could be used for optical detection of mw/rf signals by exploiting the most efficient detectors for optical photons~\cite{Bagci2014,Takeda2018,Haghighi2018,Simonsen2019a,Simonsen2019b}. 

The easiest way to bridge the enormous energy gap is to use a mediator simultaneously coupled to both mw/rf and optical modes. There have been a variety of proposals using different kinds of mediating systems,  including opto-electro-mechanical systems~\cite{Wang2012,Tian2012,Barzanjeh2012,Hill2012,Andrews2014,Higginbotham2018,Malossi2021,Brubaker2022}, atomic ensembles~\cite{Hafezi2012,Petrosyan2019,Tu2022,Kumar2023}, electro-optical systems~\cite{Tsang2010,Rueda2019,McKenna2020,Hease2020,Sahu2022,Sahu2023} and~magnons~\cite{Hisatomi2016}.

In opto-electro-mechanical platforms, a~mechanical resonator is coupled to mw/rf photons either capacitively~\cite{Bagci2014,Takeda2018,Haghighi2018,Simonsen2019a,Simonsen2019b,Andrews2014,Higginbotham2018,Malossi2021,Brubaker2022} or~via the piezoelectric effect~\cite{Vainsencher2016,Balram2016,Schneider2019,Stockill2022,Jiang2019,Shao2019,Jiang2020,Han2020,Forsch2020,Mirhosseini2020} and~dispersively via radiation pressure with the optical mode(s). Here, we shall focus on these platforms, which present a high degree of flexibility. In~fact, transduction occurs at motional sidebands of the microwave and optical drivings and, depending on the detunings, one can exploit either direct transduction with a beam-splitter-like interaction~\cite{Wang2012,Tian2012,Hill2012,Andrews2014,Higginbotham2018} or~state transfer via quantum teleportation enabled by the microwave-optical entanglement mediated by the mechanical mode~\cite{Barzanjeh2012,Zhong2020}. Moreover, by~including a second mechanical mode~\cite{Piergentili2018aa,Piergentili2021tn,Piergentili2021wg} and by properly driving both mw/rf and optical modes with multiple tones in a phase-controlled way, one can realize nonreciprocal photon transmission/conversion~\cite{Metelmann2015,Xu2016,Bernier2017,Barzanjeh2017,Miri2017,Shen2018,Ruesink2018,Malz2018,Mercier2020,Eshaqi2022}. In~fact, the~two mechanical modes establish two distinct paths of photon transmission which may destructively interfere, breaking the symmetry between the two directions. This has enabled the realization of isolators, circulators and directional amplifiers both at microwave~\cite{Bernier2017,Barzanjeh2017,Mercier2020} and at optical~\cite{Shen2018,Ruesink2018} frequencies, while nonreciprocal microwave-optical transducers proposed in~\cite{Xu2016,Eshaqi2022} have not been experimentally realized yet.
Typical figures of merit in quantum transduction are the conversion efficiency (or the fidelity of the transferred quantum state), the~added noise and~the bandwidth~\cite{Zeuthen2020}. The~two latter quantities are relevant also when transduction is exploited for sensing, as~it occurs in recently demonstrated opto-electro-mechanical devices realizing optical preamplification and detection of rf signals in magnetic resonance imaging~\cite{Simonsen2019a,Simonsen2019b}. In~these devices, the~detection bandwidth of rf signals could be enhanced and engineered by again exploiting  the interference between the optomechanical interaction pathways mediated by two different mechanical modes~\cite{Haghighi2018}. 

Various solutions have been adopted for the explicit design of opto-electro-mechanical transducers. In~some cases,  transducers are fabricated fully from a piezoelectric material, including the optomechanical cavity~\cite{Vainsencher2016,Balram2016,Schneider2019,Jiang2020, Stockill2022}. Other approaches exploit hybrid solutions where the optomechanical part is fabricated from Si, for~example, while coupling to the microwave input is through an added piezoelectric resonator~\cite{Jiang2019,Shao2019,Han2020,Forsch2020,Mirhosseini2020}. In~these cases, quasi-resonant direct interaction between the mechanical resonator and a microwave resonator at GHz frequencies is exploited for~transduction.

Electro-mechanical capacitive coupling is instead used in  cases of metalized membranes~\cite{Bagci2014,Takeda2018,Haghighi2018,Simonsen2019a,Simonsen2019b,Andrews2014,Higginbotham2018,Malossi2021,Brubaker2022} which, with~an in-front electrode, may be placed within an optical Fabry--Perot cavity for radiation--pressure coupling. 
The metalized membrane and the electrode form a capacitor of an LC resonator whose rf photons are modulated by the membrane motion. This membrane-based geometry is  used in two cases. (i) rf-optical transduction~\cite{Bagci2014,Takeda2018,Haghighi2018,Simonsen2019a,Simonsen2019b,Malossi2021}, in~which the vibrational mode of the membrane and the LC resonator are quasi-resonant and~the working point is set by an applied DC voltage bias. (ii) mw-optical transduction~\cite{Andrews2014,Higginbotham2018,Brubaker2022} between the fields at mechanical sideband frequencies with respect to the corresponding optical and microwave driving~fields. 

Here, we focus on quasi-resonant rf-optical transduction based on metalized membranes and~we report a novel design of an opto-electro-mechanical modulator (OEMM) based on a resonating silicon nitride nanomembrane, which exhibits room-temperature $Q$-factors of $10^6$ in the MHz range. Square-shaped metalized nanomembranes with the LC circuit in FR4 and quartz substrates were presented in other works with quality factors $10^5$~\cite{Bagci2014,Haghighi2018} and resulting in more added thermal noise in the transduction scheme. 
Our device can be optimized to be embedded into the optical cavity for the sympathetic ground state cooling of an LC resonating circuit as proposed in~\cite{Malossi2021}. We show the fundamental properties of this novel device and the fabrication~process. 

The paper is organised as follows. In~Section~\ref{sec:design}, we describe the design strategy we have followed, while in Section~\ref{sec:fabrication} we provide the details of the fabrication process employed for the realization of the OEMM device. In~Section~\ref{sec:characterization}, we provide a characterization of the electro-mechanical properties and performance of the device and~in Section~\ref{sec:discussion} we summarize the results and discuss the perspectives, the~future research direction and~potential~applications. 

\section{Design Strategy of the~OEMM} \label{sec:design}
\begin{figure}[H]
\includegraphics[width=12.5cm]{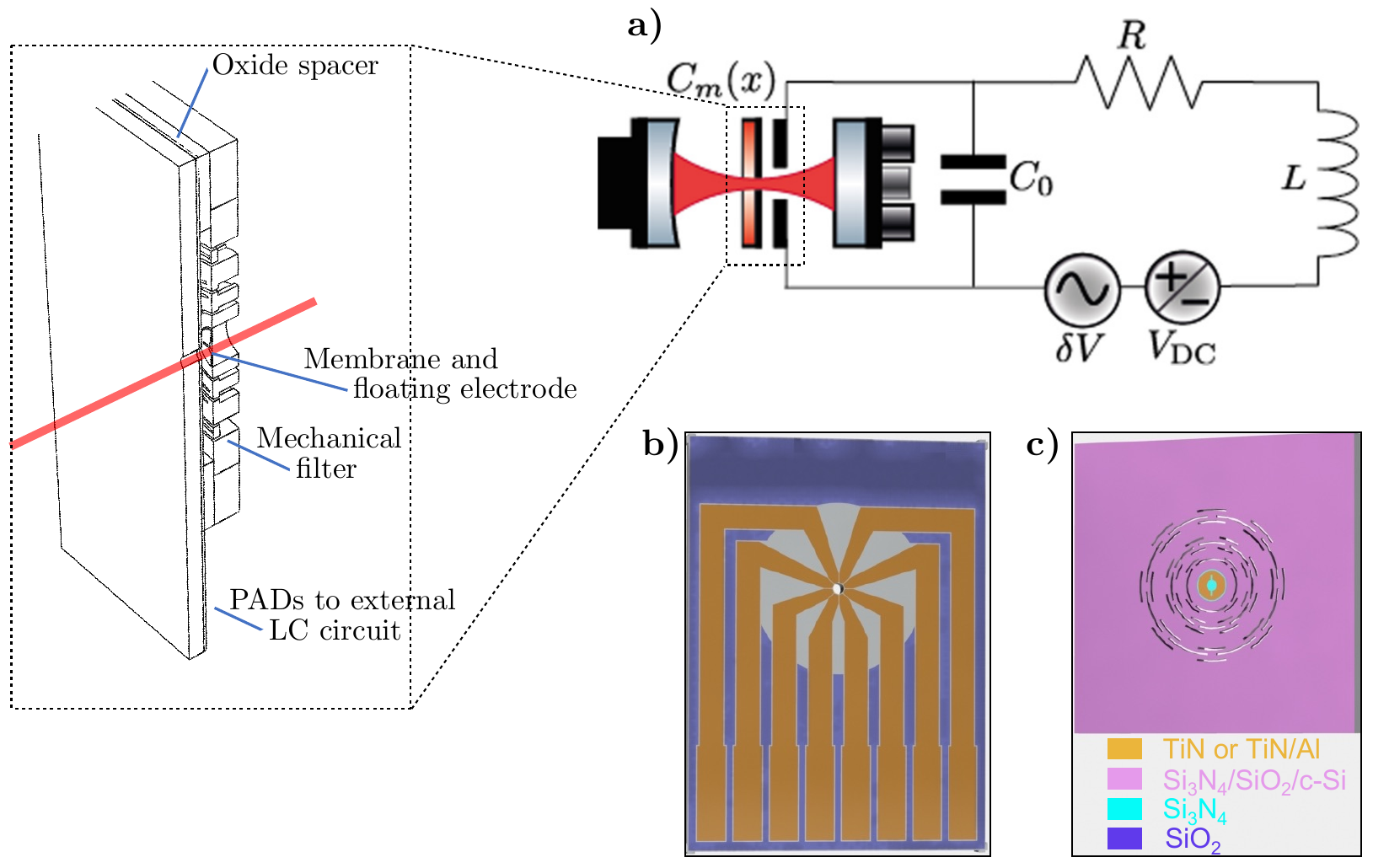}
\caption{Opto--Electro-Mechanical Modulator.(\textbf{a}) Schematic of the setup with a detailed view of the cross-section of the OEMM embedded in the high-finesse Fabry--Perot Cavity. The~rf weak $\delta V$ signal is transferred to the cavity optical field after polarizing the coupling capacitor $C_m(x)$ with a DC voltage bias $V_{DC}$. (\textbf{b},\textbf{c}) The two components of the OEMM device:  the electrode connected to the LC circuit (\textbf{b}) and the floating electrode on the silicon nitride membrane (\textbf{c}). Materials are also specified, while the substrate is floating-zone crystalline silicon for both~components.\label{lab:fig1}}
\end{figure}

The device comprises a metal-coated silicon nitride nanomembrane that is coupled via radiation pressure to a cavity field and~capacitively coupled to an rf resonant circuit via the position-dependent capacitance, as~shown in Figure~\ref{lab:fig1}. In~our design, we need to keep the thermal noise low by providing the device with an on-chip mechanical filter for shielding the substrate's recoil losses~\cite{Borrielli2016}. In~particular, the~membrane is clamped to a hollow silicon cylinder connected with flexural joints to the two-stage shield shown in Figure~\ref{lab:fig1}c. The~mechanical filter and the thick hollow cylinder cause the dissipation of the membrane to be dominated only by the intrinsic losses of the two films. The~high stress in the membrane dilutes the loss of both films as discussed below. Moreover, the~intrinsic losses of a highly stressed circular membrane are related to the dilution parameter $\lambda$, which can be expressed in terms of the aspect ratio $h/R$, where $R$ is the radius and $h$ the thickness, as~$\lambda=\frac{h}{R}\sqrt{\frac{Y}{12(1-\nu^2)\sigma_0} }$, where $\sigma_0$ is the bi-axial intrinsic stress, $Y$ the Young's modulus and~$\nu$ the Poisson's ratio. The~dilution parameter can be equivalently defined in terms of the flexural rigidity $D$ of a film of a given material as  $\lambda=D/(\sigma_0 h R^2)$.
The intrinsic losses can be decomposed into the edge and distributed losses~\cite{Serra2021,Yu2012}, which depend upon the $\lambda$ parameter linearly and quadratically, respectively. 

Edge loss is due to the high curvature  variations in the resonating modes in the clamping region which extends about fivefold the thickness of the SiN membrane ($5 \times h$) and affects the quality factor of the low-frequency modes. While the contribution to this loss of metal TiN coating can be easily avoided via a selective patterning of the metal film, the~distributed loss degrades the overall quality factor according to the weighted sum of the bending dissipation of the two films~\cite{Vengallatore2011}. In~the following, we demonstrate that the distributed loss in the TiN coating is diluted by the tensile stress $\sigma_0$ in the SiN membrane. Starting from the definition of the quality factor of a bilayer (SiN/TiN) due to the internal bending of the films, we can write:
\begin{equation}
Q_{SiN/TiN}= \frac{2 \pi W_{SiN}^t}{\Delta W_{SiN}+\Delta W_{TiN}}=\left (\frac{W_{SiN}^b}{W_{SiN}^t} Q_{SiN}^{-1}+\frac{W_{TiN}^b}{W_{SiN}^t} Q_{TiN}^{-1}\right )^{-1}\,,
\label{eq:Qbilayer}
\end{equation}
where  $\Delta W_{SiN},\Delta W_{TiN}$ are the dissipated energies due to bending, while $W_{SiN}^b,W_{TiN}^b$ are the stored bending energies of the silicon nitride film and the titanium nitride films, respectively. ~$W_{SiN}^t$ is the total tensile stored energy in the SiN film. In~addition, the~$ Q_{SiN}^{-1}, Q_{TiN}^{-1}$ are the intrinsic losses of the silicon nitride layer and the titanium nitride in cases of zero stress, respectively. For~a circular membrane, the~energy terms can be evaluated with the following formulas: 
\begin{equation}
W_{SiN}^b=\frac{Y_{SiN}}{(1-\nu_{SiN}^2)} \int_0^{2 \pi} d\theta \int_{-\frac{h_{SiN}}{2}}^\frac{h_{SiN}}{2} z^2 dz \int_0^R \left(\frac{\partial^2 u(r)}{\partial^2 r} +\frac{1}{r} \frac{\partial u(r)}{\partial r}\right)^2 r dr\,,
\label{eq:WSiNb}
\end{equation}
\begin{equation}
W_{TiN}^b=\frac{Y_{TiN}}{(1-\nu_{TiN}^2)} \int_0^{2 \pi} d\theta \int_{\frac{h_{SiN}}{2}}^{\frac{h_{SiN}}{2}+h_{TiN}} z^2 dz \int_{R_i}^{R_e}\left(\frac{\partial^2 u(r)}{\partial^2 r} +\frac{1}{r} \frac{\partial u(r)}{\partial r}\right)^2 r dr\,,
\label{eq:WTINb}
\end{equation}
and the tensile energy is: 
\begin{eqnarray}
W_{SiN}^t= \int_0^{2 \pi} d\theta \int_0^R  \frac{h_{SiN} \sigma_0}{2}\left (\frac{\partial u(r)}{\partial r}\right)^2 r dr\,.
\label{eq:tensEnergy}
\end{eqnarray}
Using the equation of vibrations of axisymmetric modes of a membrane under uniform tensile stress $\sigma_0$ 
\begin{equation}
\nabla_r u(r)=-\frac{\rho \omega^2}{\sigma_0} u(r)=-\Omega^2 u(r)\,,
\end{equation}
with $\rho$ being the membrane's density, we derive the axisymmetric solution $u_{n0}(r)=C_{0n} J_0(\alpha_{0n}r/R)$ in cases of clamped--clamped boundary conditions, where $ J_0$ is the zero-th order Bessel function of the first kind, $\alpha_{0n}$ are its zeros and~$\Omega_{0n}=\alpha_{0n}/R$. According to this definition, the~energy terms in Equations~(\ref{eq:WSiNb}) and (\ref{eq:WTINb}) transform as
\begin{align}
W_{SiN}^b=&\,\frac{Y_{SiN}}{(1-\nu_{SiN}^2)} \int_0^{2 \pi} d\theta \int_{-\frac{h_{SiN}}{2}}^\frac{h_{SiN}}{2} z^2 dz \int_0^R \Omega_{0n}^4 C_{0n}^2 J_0^2(\alpha_{0n}r/R) r dr \nonumber\\
= &\,\pi R^2 \Omega_{0n}^4 C_{0n}^2 D_{SiN} J_1^2(\alpha_{n0}),
\end{align}
\begin{align}
W_{TiN}^b=&\,\frac{Y_{TiN}}{(1-\nu_{TiN}^2)} \int_0^{2 \pi} d\theta \int_{\frac{h_{SiN}}{2}}^{\frac{h_{SiN}}{2}+h_{TiN}} z^2 dz \int_{R_i}^{R_e} \Omega_{0n}^4 C_{0n}^2 J_0(\alpha_{0n}r/R)^2 r dr \nonumber\\ 
=&\, \pi R^2 \Omega_{0n}^4 C_{0n}^2 D_{TiN}
J_1(\alpha_{n0})^2 \left[1-\frac{R_i^2}{R^2}\left(\frac{J_0(\alpha_{0n} R_i/R)^2}{J_1(\alpha_{0n})^2}+
\frac{J_1(\alpha_{0n} R_i/R)^2}{J_1(\alpha_{0n})^2}\right)\right]\,,
\end{align}
where $R_i, R_e$ are the internal and external radius of the TiN metal coating and~$ J_1$ is the first-order Bessel function of the first kind. In~the above equation, without~loss of generality, we assume that $R_e \simeq R$.  In~the fabricated  device,  the condition to avoid 
edge loss is $R_e < R$. 

We now consider the tensile stored energy, which can be written starting from Equation~(\ref{eq:tensEnergy}) as: 
\begin{eqnarray}
W_{SiN}^t= \pi h_{SiN} \sigma_0 \frac{\alpha_{0n}^2 C_{0n}^2 J_1(\alpha_{0n})^2}{2}\,.
\end{eqnarray}
Using the equation above and~the definition of the dilution parameter for the SiN and the TiN films 
given above~\cite{Serra2021}, we can rewrite Equation~(\ref{eq:Qbilayer}) as 
\begin{eqnarray}
Q_{SiN/TiN}=\left(\alpha_{0n}^2 \lambda_{SiN}^2 Q_{SiN}^{-1}+\alpha_{0n}^2 \lambda_{TiN}^2 [1-f(R_i)] Q_{TiN}^{-1}\right)^{-1}\,,
\label{eq:Qbilayer1}
\end{eqnarray}
where:
\begin{eqnarray}
f(R_i)=\frac{R_i^2}{R^2}\left(\frac{J_0(\alpha_{0n} R_i/R)^2}{J_1(\alpha_{0n})^2}+
\frac{J_1(\alpha_{0n} R_i/R)^2}{J_1(\alpha_{0n})^2}\right)
\end{eqnarray}
is the function that modulates the distributed dissipation on the TiN film. 
By looking at Equation~(\ref{eq:Qbilayer1}), the~two terms represent the distributed losses for each film and~show that the loss is diluted by the stress in the SiN layer. Moreover, the~losses due to the TiN layer can be further reduced by increasing the $f(R_i)$ term.  

 The ultimate limit of the mechanical Q factor of the bilayer film is due to the edge loss ($2 \lambda_{SiN} Q_{SiN}^{-1}$) in the SiN layer that must be included in Equation~(\ref{eq:Qbilayer}):
\begin{eqnarray}
Q_{SiN/TiN}^{Tot}=\Big[(2 \lambda_{SiN}+\alpha_{0n}^2 \lambda_{SiN}^2) Q_{SiN}^{-1}+\alpha_{0n}^2 \lambda_{TiN}^2 [1-f(R_i)] Q_{TiN}^{-1}\Big]^{-1} \,.
\label{eq:QtotDis}
\end{eqnarray}

\begin{figure}[H]
\includegraphics[height=60 mm, angle=0]{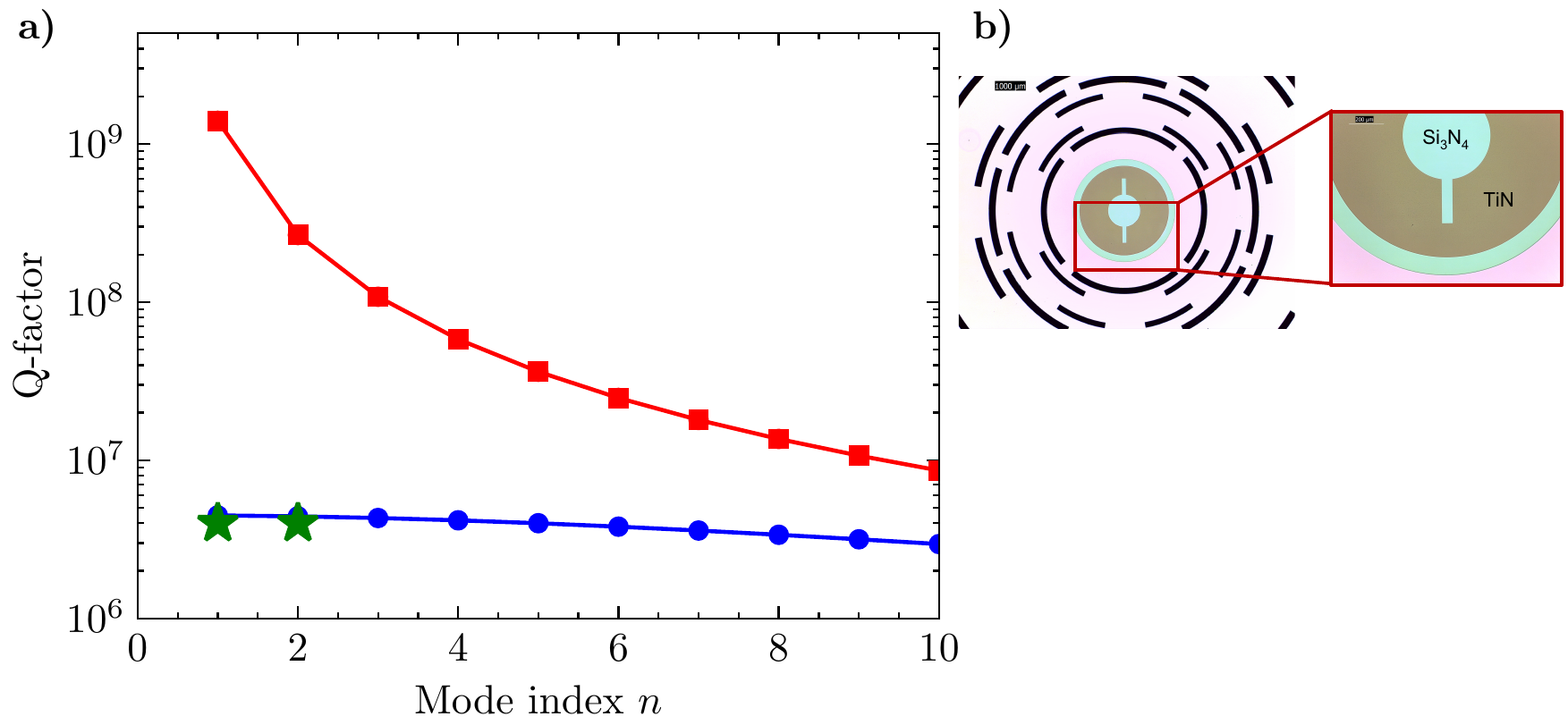}
\caption{Dissipation of the metalized membrane. (\textbf{a}) The $Q$-factor of the bilayer SiN/TiN membrane without edge losses $Q_{SiN/TiN}$ (red) and with edge loss of the SiN layer $Q_{SiN/TiN}^{Tot}$ (blue). Green stars are the experimental points of the measured $Q$-factor for the two first axisymmetric modes with index (0,1) and (0,2).
(\textbf{b}) Optical image of the SiN stoichiometric nanomembrane (light blue) with the TiN layer (brown). The~membrane is endowed with the on-chip shield for recoil losses (see Ref.~\protect~\cite{Borrielli2016}) (right). Detailed  view of the TiN notch in the membrane electrode used for the mode identification and the component assembly.}
\label{fig5}
\end{figure}

To the authors' knowledge, there are no data for evaluating the internal friction of the TiN coating; hence,  a precise estimation of the overall $Q$-factor would be fair. We assume an intrinsic loss of TiN equal to $10^{-3}$ as usually done for simulating the effects of the metal dissipative layer. Results concerning the overall dissipation and the distribution are shown in Figure~\ref{fig5}{a} for the first ten asymmetric modes by using the values reported in Table~\ref{tab1}. The~above relations demonstrate that the thickness and the shape of the metal coating on a stressed membrane can be arbitrary if patterning starts from a few microns of the membrane's edge. The~distributed losses of the two films are always lower than the edge loss in the SiN layer.  Similar behavior was experimentally observed in~\cite{Yu2012} where a hybrid SiN membrane with an aluminum layer was presented. The~detrimental effect of the intrinsic loss of aluminum film was observed in a full-coated version of the~device. 

The eight-segment electrode (see Figure~\ref{lab:fig1}{b})  has been designed to satisfy the condition of constructive interference between the mechanical modes that improve the transduction of rf signals into the optical output within the frequency band between the two mechanical resonances.
As described in~\cite{Haghighi2018}, this modifies the electromagnetic coupling coefficients through the effective areas $A_{eff}$ that are determined by the partial capacitance between the floating electrode in the membrane and the TiN/Al paths in the fixed electrode. Hence, the~overall variable capacitance between two electrodes (+,$-$) in front of the membrane can be obtained as:
\begin{eqnarray}
C_m=\left(\frac{1}{C_{+}}+\frac{1}{C_{-}}\right)^{-1}\,.
\label{eq:C_m}
\end{eqnarray}
Assuming that the curvature of the membrane is sufficiently small so that we can take it to be locally flat: 
\begin{eqnarray}
C_{\pm}= \int_0^{2\pi} d\theta \int_0^R \frac{\zeta_{\pm}(r,\theta) \epsilon_0}{h_0+\delta z(r,\theta)} r dr\,,
\label{eq:C_pm}
\end{eqnarray}
where $h_0$ is the capacitor gap, $\zeta_{\pm}(r,\theta)$ is a function that equals 1 for points in the membrane plane that are metalized and overlap with the fixed positive or negative electrode and~is zero otherwise, $\delta z(r,\theta)=\sum_i \beta_i u_i(r,\theta)$ is the membrane displacement field relative to the steady-state configuration and~$\epsilon_0$ is the vacuum dielectric constant. The~eight-segment electrodes were designed to maximize the capacitance  variation $\frac{\partial C_{\pm}}{\partial \beta_i}$ for a given membrane's~eigenmode.

\begin{table}[H] 
\caption{Material data for the SiN and the TiN coating. Dissipation properties 
at room temperature. \label{tab1}}
\newcolumntype{C}{>{\centering\arraybackslash}X}
\begin{tabularx}{\textwidth}{CCC}
\toprule
& \textbf{SiN}	& \textbf{TiN} \\
\midrule
$Y$  [GPa] &	270 & 600\\
$\nu$ & 0.27 & 0.27 \\
$\rho$ [Kg/m$^3$]	& 2700 & -\\
$ h$ [nm] & 100 & 50 \\
$ R$ [$\upmu$m] & 740  &  - \\
$R_i$ [$\upmu$m] & - & 250 \\
$Q^{-1}$  & $2.0 \times 10^{-4}$ & $1.0 \times 10^{-3}$\\ 
\bottomrule
\end{tabularx}
\end{table}

\section{Fabrication of the OEMM~Module}   \label{sec:fabrication}
\subsection{The Opto-Electro-Mechanical Resonating~Part}
The opto-electro-mechanical resonator is made of a Low-Pressure Chemical Vapor Deposition (LPCVD) silicon nitride (SiN) nanomembrane (100~nm nominal thickness) with a partial overlay of an rf-sputtered Titanium Nitride (TiN) deposited at 50 $^{\circ}$C with a target thickness of 50~nm (see Figure~\ref{fig5}{b}). The~final thickness of the SiN membrane, measured with  an ellipsometer, is $(80 \pm 5)$~nm because of several  wet etching steps made by Hidrofluoridic Acid (HF) while the TiN thickness remains close to its nominal value. 
The nanomembrane is tensioned with a 1~GPa and insulated from the substrate by an on-chip shield where only intrinsic losses of SiN/TiN layers account for the dissipation, as~discussed in Section~\ref{sec:design}. The~double-stage filter is made of flexural silicon joints and masses realized on a Silicon-on-Insulator wafer with a thickness of 1~mm. To~etch silicon, we used a double-side bulk micromachining Inductive Coupled Plasma--Deep Reactive Ion Etching (ICP-DRIE) Bosch recipe. Descriptions of similar devices with the fabrication steps can be found in~\cite{Borrielli2016,Serra2021,Serra2018}.
\subsection{Microfabricated Coupling~Capacitor}
\begin{figure}[H]
\includegraphics[width=12cm]{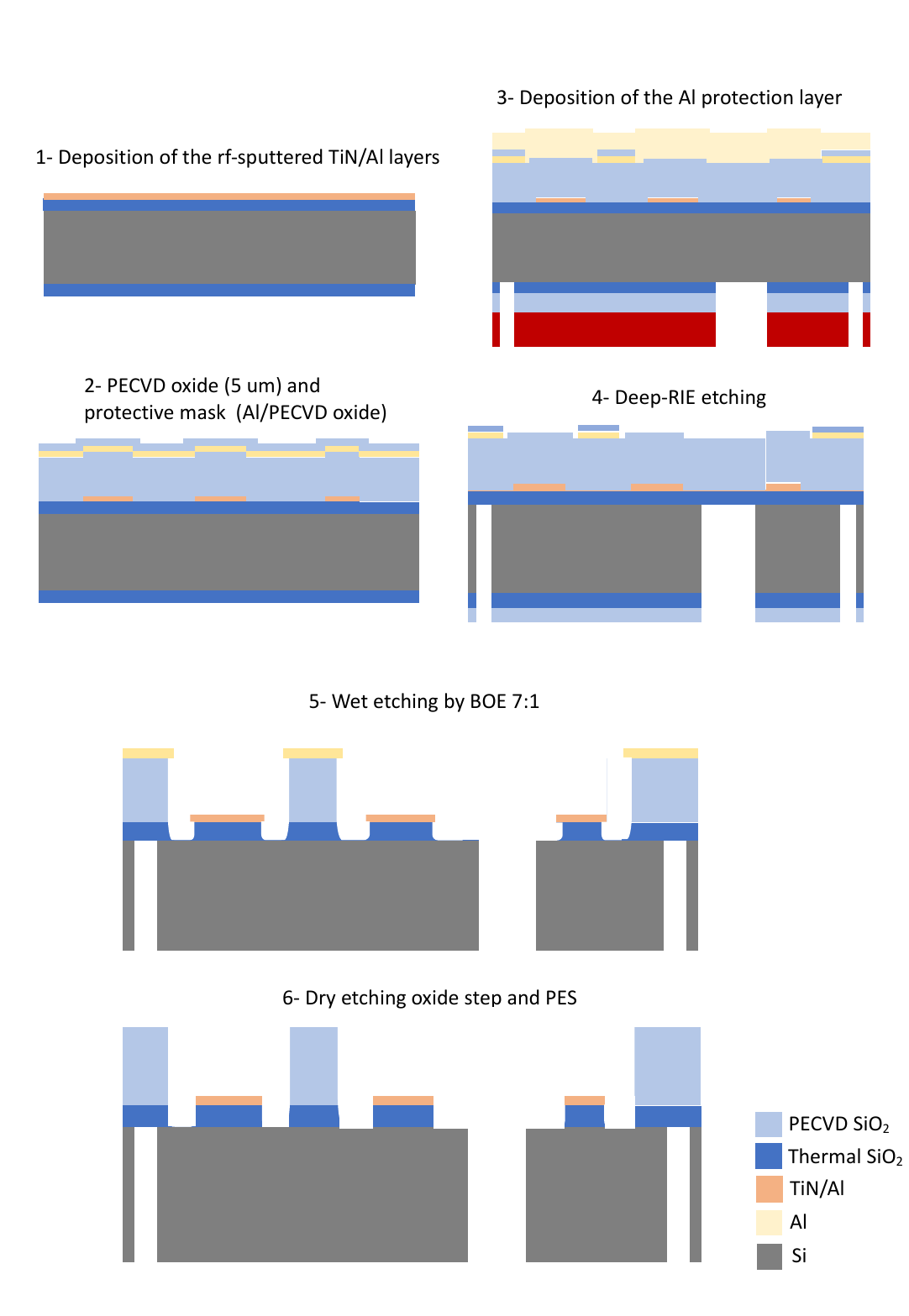}
\caption{Main steps of the microfabrication process~flow-chart. \label{fig4}}
\end{figure}

The LC circuit for the injection of the rf signal is made of an external capacitance, inductance and~a microfabricated electromechanical coupling capacitor. The~target armature distance between the membrane and the eight-segment electrode must be lower than  10~$\upmu$m. The surface/bulk-micromachining double-side microfabrication process was developed starting from a high resistivity p-type Si wafer with <100> orientation,  100~mm diameter and~thickness of 1~mm. The~principal process steps are shown in Figure~\ref{fig4}.

Fabrication starts by thermally growing 500~nm thick oxide for the electrical insulation of the metal paths. The~TiN/Al stacks were rf-sputtered with target thicknesses of 50~nm/500~nm, respectively. A~plasma etching of the TiN/Al stack was done by using an inductively coupled plasma machine based on $\rm Ar/CHF_3$ chemistry and landing on the thermal oxide layer
(see steps 1--2 in Figure~\ref{fig4}).

Deposition of 5~$\upmu$m thick PECVD TEOS oxide on top of the Si substrate that covers the metal paths was used as the etch stop layer for the ICP-DRIE etching. The~through-holes have a target diameter of 500~$\upmu$m. The~PECVD TEOS oxide works also as a spacer for the OEMM component assembly. The~oxide spacer was patterned to open the interaction area (corresponding to the filter region) and the trenches between the eight-segment electrode.  This step was done using an extra mask layer made of Al/PECVD oxide on the top of the substrate (see steps 2--3 in Figure~\ref{fig4}).  After~this step, an~aluminum capping layer was deposited on the top side for thermalization for the through-wafer silicon etching step. The~bottom side was prepared with 4~$\upmu$m PECVD oxide for the Si etching.  Deep-RIE Bosch plasma etchings with $\rm SF_6$ and $\rm C_4F_8$ as etching and passivation gasses, respectively, were done in an inductive plasma machine for the definition of the through-hole and the dicing lines.  Each device is supported on the four corners by silicon beams (width 500~$\upmu$m) connecting the device with the wafer for subsequent manual dicing (see steps in 3--4 Figure~\ref{fig4}).  

A BOE (buffered oxide etch) 1:7 was carried out on each device to remove the PECVD oxide (see step 5 in Figure~\ref{fig4}) around the Al mask and TiN/Al layer. The~first mask protects the PECVD oxide spacer, and the second the insulation thermal oxide. 
After BOE 1:7, the~ Al stripping was done with PES-77-19-04, which mainly consists of phosphoric acid ($\rm H_3PO_4$).
In fact, PES can be a mixture 1--5$\%$ $\rm HNO_3$ (for Al oxidation), 65--75\% $\rm H_3PO_4$ (to dissolve 
the $\rm Al_2O_3$),  5--10$\% $  $\rm CH_3COOH$ (for wetting and buffering) and~$\rm H_2O$ dilution to define the etch rate at a given temperature. There is a <1~$\upmu$m undercut on the sidewall of the $\rm TiN/Al/SiO_2$ oxide.  The~minimum distance between the metal paths and the oxide spacer is 50~$\upmu$m while the distance from the central hole is 30~$\upmu$m.  Dry etching was performed with a plasma machine to remove unwanted extra oxide layers between the metal paths exploiting the selectivity of the TiN metallization with respect to oxide  (see step 6 in Figure~\ref{fig4}). The~TiN layer was used also because it is inerted in most chemical baths and simplifies microfabrication. The~PECVD oxide spacer has a final thickness of 1.1~$\upmu$m. The~cleaning was done with an ultrasonic bath and a custom Teflon holder inside a glass beaker heated by a hot plate. Devices are positioned to facilitate the bubbling of hot 60~$^{\circ}$C DI water and a pipe for the continuous replacement of the exhausted DI water. 
\subsection{OEMM Assembly~Procedure}
\begin{figure}[H]
\includegraphics[height=35 mm, angle=0]{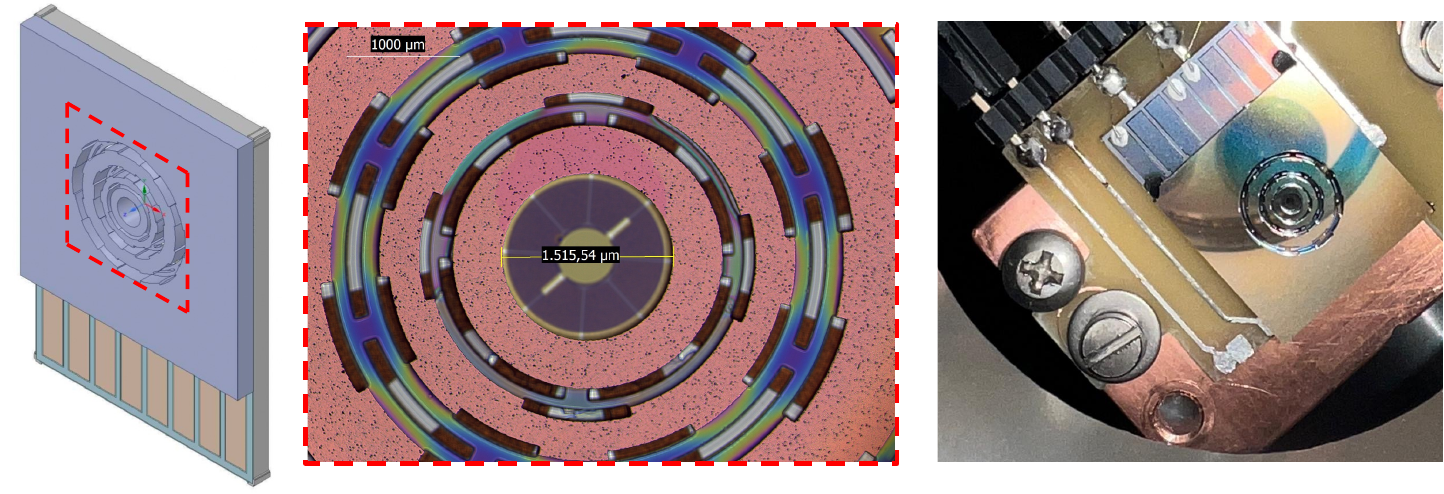}
\caption{Detailed view of the interaction region in the OEMM assembled device (\textbf{left}). The~OEMMs clamped to the OFHC copper block (\textbf{center}) and~the diving PCB board used in the optical setup~(\textbf{right}).}
\label{lab:MDPI_Entropy_3}
\end{figure}

The alignment is performed on a custom designed Teflon support in which references have been integrated in the eight-segment chip. The~alignment system is mounted on a crystallographic microscope and centering is done manually using as a reference the notch on the TiN electrode on the membrane and the space between two metal tracks in the fixed electrode chip. A~membrane/electrode alignment precision on the plane of less than 35~$\upmu$m in Figure~\ref{lab:MDPI_Entropy_3} was estimated, which was in line with that of machines performing chip-to-chip bonding. The~two chips are glued by the Stycast 2850 FT glue for cryogenic applications and~applied to three device edges. Weights are placed on top of the upper chip to avoid that the glue spills into the internal region of the device, increasing the capacitor gap. The~assembled OEMM device was finally glued to an oxygen-free high thermal conductivity (OFHC) copper plate for thermalization and~connected to the PCB auxiliary element with Ag paste; see Figure~\ref{lab:MDPI_Entropy_3} (right). 

\section{Characterization}  \label{sec:characterization}
\subsection{Modal Analysis of the~OEMMs}
\begin{figure}[H]
\includegraphics[height=70 mm, angle=0]{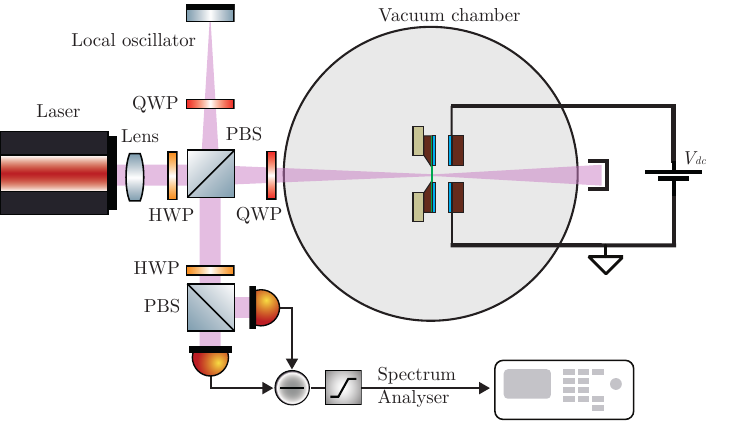}
\caption{Experimental setup. Michelson interferometer with shot-noise limited homodyne~detection.}
\label{fig:fig5setup}
\end{figure}

\begin{figure}[H]
\includegraphics[height=50 mm, angle=0]{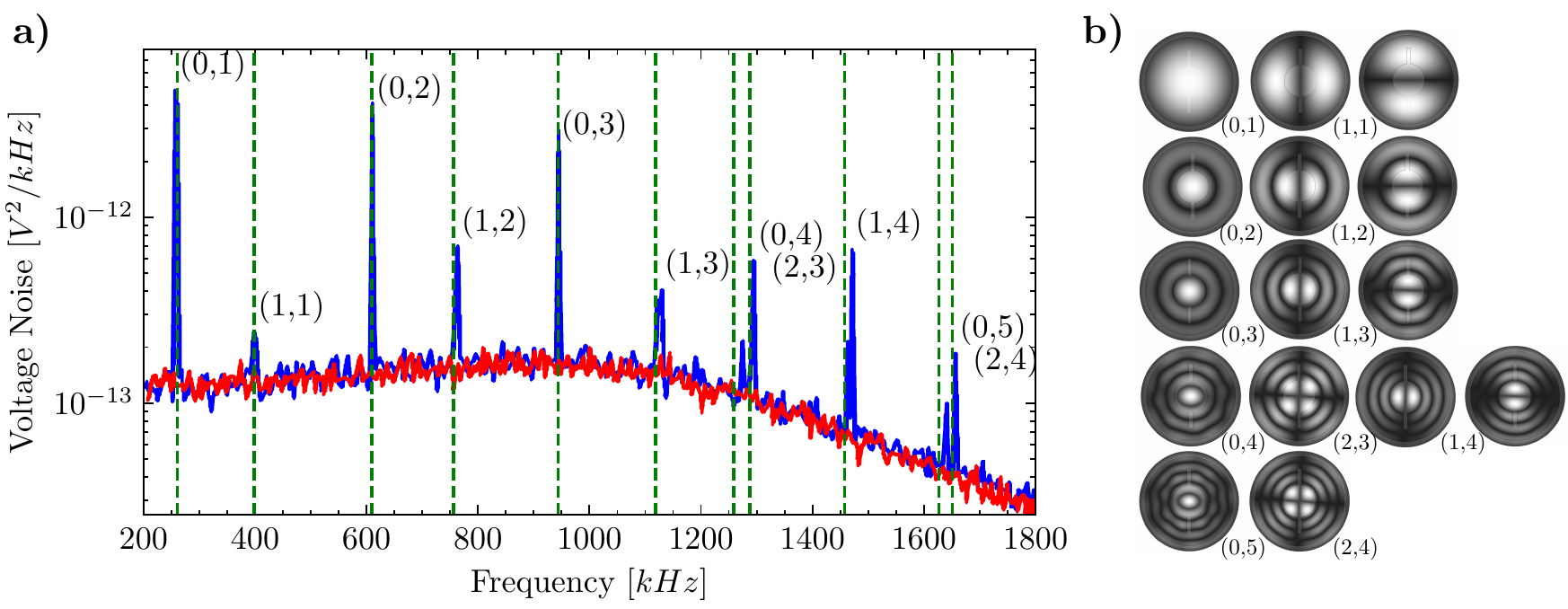}
\caption{Mechanical mode characterization. (\textbf{a}) Voltage spectral noise of the homodyne signal (blue) and the shot-noise contribution (red). Mechanical modes of the functionalized Si$_3$N$_4$ membrane correspondence to the peaks of the spectrum. The~first mode has frequency $f_{(0,1)}=260.65$ kHz. The~green dashed lines are the calculated frequencies from the FEM simulation. (\textbf{b}) Calculated mode shapes via FEM simulation, corresponding to the calculated mode frequencies present in the spectrum. Axisymmetric modes and modes with two-fold degeneracy are classified according to the number of nodal and circumference indexes. Frequency 
increases from the top to the bottom and from left to~right.}
\label{fig:FigModes}
\end{figure}

The  modal frequencies and~ the quality factors of the mechanical oscillator are measured by means of an interferometric technique with shot-noise limited homodyne detection. The~setup is shown in Figure~\ref{fig:fig5setup} and~the measurement techniques are discussed in~\cite{Haghighi2018}.   
In Figure~\ref{fig:FigModes}{a} the voltage noise spectrum of the homodyne signal at the output of the interferometer is presented. The~resonances are related to the peaks of the voltage noise spectrum emerging from the background shot noise.

The modal frequencies and~modal indexes are derived from a Finite Element Method (FEM) simulation of the eigenfrequencies of a SiN membrane with the TiN layer under tensile stress. Data used in the simulation are presented  in Table~\ref{tab1}. We note that the relative error $R_E$ between the experimental results and those of the simulation is always below 3$\%$, as~reported in Table~\ref{tab2}.
In Figure~\ref{fig:FigModes}{b} the simulated modal shape functions of the first nine modes are reported. 
The degenerate modes are split in frequency depending on the fabrication irregularities (hole eccentricity) and~the presence of the TiN layer. 
Modes with moving mass in the central hole of the membrane are close to those of circular membranes, but~the two-fold degeneracy is split in~frequency.

\begin{figure}[H]
\includegraphics[height=65 mm, angle=0]{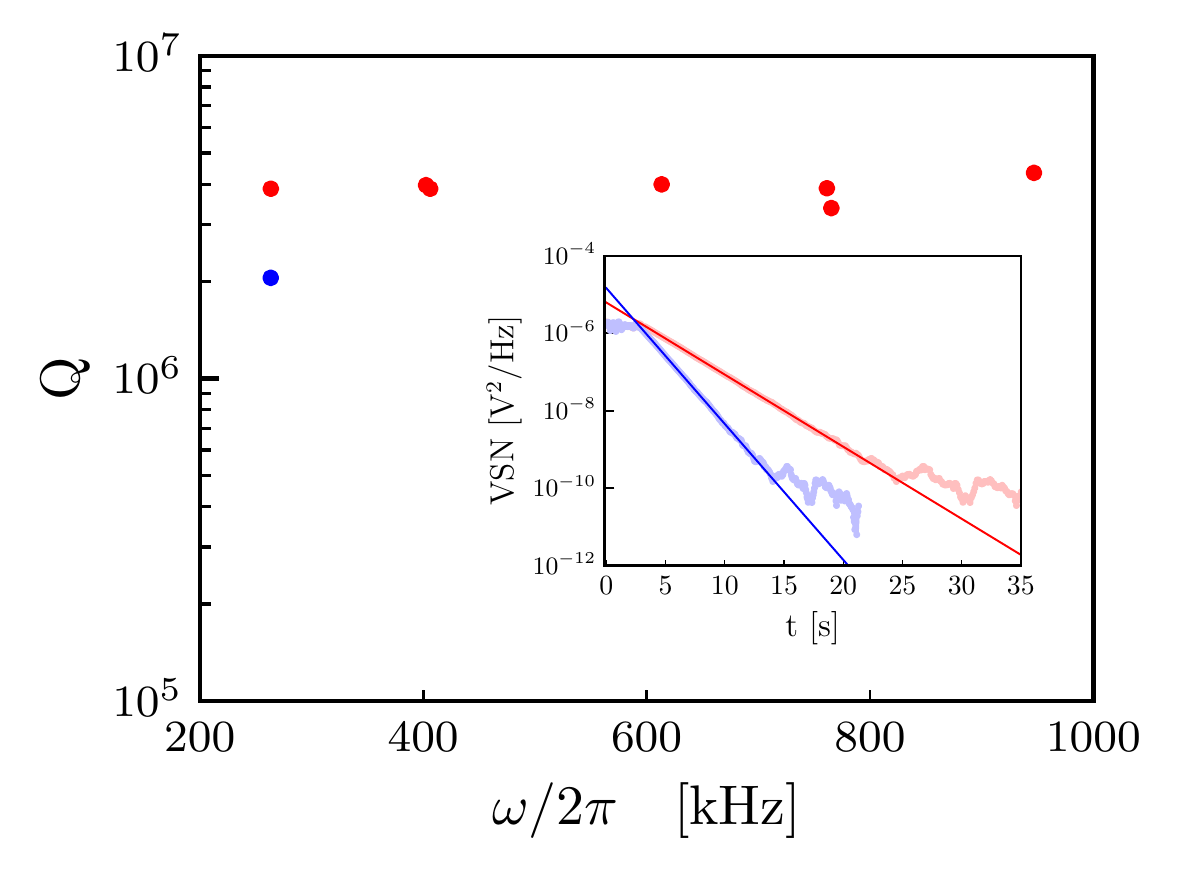}
\caption{Measurement of the mechanical $Q$-factor for different modes after the final assembling of the device. Red points correspond to measurements at $V_{DC}$ = 0~V, the~blue point corresponds to the measurement at $V_{DC}$ = 30~V. Inset: voltage spectral noise (VSN) density of the homodyne signal during the ring-down measurement of the fundamental mode (0,1) at $V_{DC}$ = 0~V (dotted red line) and $V_{DC}$ = 30~V (dotted blue line) and~corresponding fit (continuous lines), with~best fit values $\tau_{0V} = 4.668 \pm 0.008$~s and $\tau_{30V} = 2.476 \pm 0.004$~s, respectively.}
\label{lab:Fig4}
\end{figure}

\unskip
The characterization of the quality factor was performed by the ring-down technique. The~mechanical oscillator is excited on-resonance by a piezoelectric element and then left to relax in order to measure the amplitude decay time.
In Figure~\ref{lab:Fig4}, we present the mechanical $Q$-factor measurements for the assembled device in the frequency range [200~kHz, 1~MHz], while in the inset we present the ring-down measurement for the mode (0,1). The~ring down measurement signal is given by the voltage spectral noise density (VSN) of the homodyne detection signal, measured by a spectrum analyzer, during~the excitation--dexcitation cycle of the mechanical oscillator mode.
The overall $Q$-factor at room temperature after the assembling procedure is still of the order of $10^6$, which is compatible for the foreseen application of the system~\cite{Malossi2021}.

\begin{table}[H] 
\caption{Measured and computed FEM of the detected eigenfrequencies of the OEMM membrane resonator. The~corresponding shape functions are represented in Figure~\ref{fig:FigModes}{b}.}
\newcolumntype{C}{>{\centering\arraybackslash}X}
\begin{tabularx}{\textwidth}{CCCC}
\toprule
\textbf{Mode Index} & \boldmath{$f^{FEM}$} \textbf{[kHz]}	& \boldmath{$f^{meas}$} \textbf{[kHz]} & \boldmath{$|R_E| \%$}\\
\midrule
(0,1) & 260.645	 & 258.786 & 0.718\\
(1,1) & 398.802   & 399.587 & 0.196 \\
(0,2) & 609.854   & 611.659 & 0.295\\
(1,2) & 755.813   & 764.629 & 1.153 \\
(0,3) & 943.985   & 943.094 & 0.094\\
(1,3) & 1118.259   & 1129.09 & 0.959\\
(0,4) & 1258.374   & 1296.55 & 2.944\\
(2,3) & 1288.075   & 1275.69 & 0.971 \\
(1,4) & 1457.844   & 1471.24 & 0.910\\
(0,5) & 1626.986    & 1640.15 & 0.802 \\
(2,4) &  1651.324  & 1658.11 & 0.409\\
\bottomrule
\end{tabularx}
\label{tab2}
\end{table}\unskip


We also investigate the effect of the electro-mechanical coupling on the quality factor of the mechanical resonator. In~Figure~\ref{lab:Fig4}, we present the measurement of the quality factor of the mode (0,1) at $V_{dc}=30$ V (blue dot) via the ring-down technique (inset). The~measured decrease in the quality factor may be connected to the membrane tension change due to the presence of the electrostatic force and~a further systematic investigation of this effect is planned in future~developments.



\subsection{Electro–Mechanical Characterization of OEMM~Device}

In order to characterize the electromechanical properties of the device, we now consider the coupling effect induced by a DC voltage bias, $V_{DC}$, applied to the device electrodes, which induces a linear capacitive coupling between the electrical and mechanical system~\cite{Bagci2014, Haghighi2018, Malossi2021}. The~electro-mechanical coupling induces a static shift in the mechanical frequency, caused by the electrostatic force, arising between the metalized membrane surface and the electrodes, like the force arising between the plates of a capacitor. For~a capacitively-coupled system, the~angular frequency shift is given by~\cite{Bagci2014}: 
\begin{equation}
\Delta\Omega_m =- \frac{d^2 C(x)}{dx^2} \frac{V_{DC}^2}{2 m \Omega_m} 
\end{equation}
where $C(x)$ is the capacitance of the electrode/membrane capacitor, $x$ is the position of the mechanical oscillator (the membrane), $m$ is the mass of the oscillator, $\Omega_m$ is the unperturbed mechanical angular frequency and~$V_{DC}$ is the applied voltage.
As shown in~\cite{Haghighi2018, Malossi2021}, the~frequency shift can be written as a function of the geometrical parameter of the device as follows:
\begin{equation} 
\Delta f_m =-\frac{\epsilon_0}{8\pi^2}\frac{A_{eff}}{m_{eff}d^3 f_0} V^2_{DC}\,,
\label{eq:fitshif}
\end{equation}
where $\epsilon_0$ is the vacuum electrical permittivity, $d$ the average distance between the electrodes and the membrane, $A_{eff}$ the overlapping area between the electrode and the membrane, weighted by the mechanical mode shape, $m_{eff}$ the mechanical oscillator mass and~$f_0$  the unperturbed mechanical oscillator mode~frequency.

\begin{figure}[H]
\includegraphics[height=60 mm, angle=0]{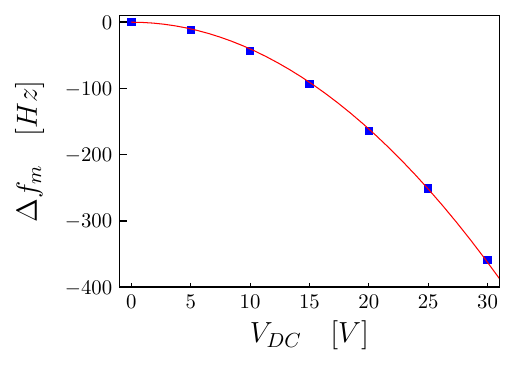}
\caption{Measurement 
 of the frequency shift of the (1,1) mode as a function of the DC voltage bias $V_{DC}$ (blue square). The~red line is the fit of the data  using Equation~(\ref{eq:fitshif}), with~best value of the average distance between the electrodes and the membrane of $d=(5.12\pm 0.14)$~$\upmu$m and~ using the following parameters: effective area $A_{eff} = 0.075$~mm$^2$; membrane mass $m_{eff} = 420$~ng;  measured unperturbed mode frequency $f_0=f_{11} = 399587$~Hz.}
\label{lab:Fig6}
\end{figure}

%
In Figure~\ref{lab:Fig6}, we show the measurement of the frequency shift of the mode (1,1) as a function of the voltage $V_{DC}$, which is applied to two electrodes that are symmetric with respect to the nodal axis of the (1,1) mode. We estimated the average distance between the electrodes and the membrane to be $d=(5.12\pm 0.14)$~$\upmu$m, by~fitting the data with Equation~(\ref{eq:fitshif}) and~by using the following parameters: effective area $A_{eff} = 0.075$~mm$^2$; membrane mass $m_{eff} = 420$~ng;  measured unperturbed mode frequency $f_0=f_{11} = 399587$~Hz.

\section{Discussion and Future~Perspectives} \label{sec:discussion}

In this work, we present a membrane-based  ultra-coherent OEMM that is potentially able to realize the bidirectional transduction of weak signals between the MHz domain into the THz (optical) domain. This kind of devices could be employed for the transduction of quantum signals at the input and output of solid-state superconducting processors, usually operated at rf/microwave frequencies, to~optical fiber connections which are suitable for long distance quantum communication~\cite{Lauk2020,Kimble2008,Pirandola2018}. The~same transduction could be useful for quantum sensing because weak rf signals could be read by means of easily available quantum limited optical photodetectors~\cite{Simonsen2019b}. This signal transduction can be applied to both discrete and continuous variable degrees of freedom of the radiation field and~therefore to different scenarios of quantum information encoding. In~this respect, the~sympathetic ground-state cooling of an LC resonator in a dilution refrigerator could be an important step for the realization of these quantum transducers~\cite{Malossi2021}. 
Innovative materials are also tested. For~instance, the~rf-sputtered TiN layers were used for the first time as electrodes in OEMMs with the aim of exploiting the superconductive properties to effectively cool the mechanical mode. 
Our OEMM design takes advantage of the fact that the $Q$-factor is independent of mounting even at the level of the membrane because the two-stage filter works as a shield for the substrate modes originating from the vibration of the coupling electrode, which is clamped to the holder in the cryogenic/vacuum chamber. This configuration is comparable to other state-of-the-art devices developed by other groups, with~a good reliability but with potentially higher mechanical $Q$-factors and~together with a wide bandwidth even in the presence of large DC~biases.

\vspace{6pt} 

\authorcontributions{Conceptualization, M.B., A.B., N.M., E.S., R.N. and D.V.; methodology, M.B., A.B., N.M., B.M., E.S., R.N., P.P. and D.V.; validation, A.B., G.D.G., N.M., B.M., R.N., P.P., P.M.S., 
 E.S.;
formal analysis, M.B., E.S., N.M., P.P. and D.V.; writing—original draft preparation, G.D.G., E.S., N.M., P.P. and D.V.; writing—review and editing, M.B., A.B., G.D.G., P.M.S., N.M., P.P. and D.V.; funding acquisition, E.S. and D.V. All authors have read and agreed to the published version of the manuscript.}

\funding{This research was funded by internal resources of the INFN tHEEOM-RD~experiment.}

\acknowledgments{We acknowledge the support of PNRR MUR project PE0000023-NQSTI (Italy).}

\begin{adjustwidth}{-\extralength}{0cm}

\reftitle{References}

\end{adjustwidth}
\end{document}